%% file: LATPSRJ1907p0602_Revised.tex
\documentclass[12pt,preprint]{aastex}
\usepackage{lineno}
\RequirePackage{lineno}

\def\ltsima{$\; \buildrel < \over \sim \;$}
\def\simlt{\lower.5ex\hbox{\ltsima}} 
\def\gtsima{$\; \buildrel > \over \sim \;$}
\def\simgt{\lower.5ex\hbox{\gtsima}} 
\def\arcsec{\hbox{$^{\prime\prime}$}}

\def\3eg{3EG~J093--3431}

\begin{document}
\linenumbers 
\title{PSR J1907+0602: A Radio-Faint Gamma-Ray Pulsar
Powering a Bright TeV Pulsar Wind Nebula} 
\input{authors2}
\begin{abstract}
We present multiwavelength studies of the 106.6~ms $\gamma$-ray pulsar
PSR J1907+06 near the TeV source MGRO J1908+06.  Timing observations
with \textit{Fermi} result in a precise position determination for the
pulsar of R.A. = 19$^\mathrm{h}$07$^\mathrm{m}$54\fs7(2), decl. =
+06$^{\circ}$02\arcmin16(2)\arcsec\ placing the pulsar firmly within
the TeV source extent, suggesting the TeV source is the pulsar wind
nebula of PSR J1907+0602.  Pulsed $\gamma$-ray emission is clearly
visible at energies from 100 MeV to above 10 GeV. The phase-averaged
power-law index in the energy range $E > 0.1$ GeV is $\Gamma= 1.76 \pm
0.05$ with an exponential cutoff energy $E_{c} = 3.6 \pm 0.5$ GeV. We
present the energy-dependent $\gamma$-ray pulsed light curve as well
as limits on off-pulse emission associated with the TeV source.  We
also report the detection of very faint (flux density of $\simeq\,3.4
\,\mu$Jy) radio pulsations with the Arecibo telescope at 1.5~GHz
having a dispersion measure DM = 82.1 $\pm$ 1.1 cm$^{-3}$pc. This
indicates a distance of $3.2\pm 0.6$ kpc and a pseudo-luminosity of
$L_{1400}\,\simeq \,0.035$ mJy kpc$^2$.  A {\em Chandra} ACIS
observation revealed an absorbed, possibly extended, compact ($\simlt
4^{\prime \prime}$) X-ray source with significant non-thermal emission
at R.A. = 19$^\mathrm{h}$07$^\mathrm{m}$54\fs76, decl. =
+06\degr02\arcmin14.6\arcsec\ with a flux of $2.3^{+0.6}_{-1.4}\times
10^{-14} {\rm erg}\, {\rm cm}^{-2} {\rm s}^{-1}$.  From archival {\em
  ASCA} observations, we place upper limits on any arcminute scale
2--10 keV X-ray emission of $\sim 1\times 10^{-13} {\rm erg}\, {\rm
  cm}^{-2} {\rm s}^{-1}$.  The implied distance to the pulsar is
compatible with that of the supernova remnant G40.5$-$0.5, located on
the far side of the TeV nebula from PSR J1907+0602, and the S74
molecular cloud on the nearer side which we discuss as potential birth
sites.

\end{abstract}
 

\section{Introduction\label{section-intro}}

The TeV source MGRO J1908+06 was discovered by the Milagro
collaboration at a median energy of 20 TeV in their survey of the
northern Galactic Plane \citep{Milagro07} with a flux $\sim 80$\% of
the Crab at these energies. It was subsequently detected in the
300~GeV -- 20~TeV range by the HESS \citep{HESS1908_2009} and VERITAS
\citep{VERITAS_1908_2008} experiments. The HESS observations show the
source HESS J1908+063 to be clearly extended, spanning
$\sim$0.3$^\circ$ of a degree on the sky with hints of
energy-dependent substructure. A decade earlier \citet{Lamb1997}
cataloged a bright source of GeV emission from the EGRET data, GeV
J1907+0557, which is positionally consistent with MGRO J1908+06. It is
near, but inconsistent with, the third EGRET catalog~\citep{3rdCat}
source 3EG J1903+0550 \citep{RRK01}. The Large Area Telescope
(LAT)~\citep{LATinstrument} aboard the \textit{Fermi Gamma-Ray Space
Telescope} has been operating in survey mode since soon after its
launch on 2008 June 11, carrying out continuous observations of the
GeV sky.  The \textit{Fermi} Bright Source List \citep{BSL}, based on
3 months of survey data, contains 0FGL J1907.5+0602 which is
coincident with GeV J1907+0557. The 3EG J1903+0550 source location
confidence contour stretches between 0FGL J1907.5+0602 and the nearby
source 0FGL J1900.0+0356, suggesting it was a conflation of the two
sources.

The \textit{Fermi} LAT collaboration recently reported the discovery of 16
previously-unknown pulsars by using a time differencing technique on
the LAT photon data above 300~MeV \citep{BSP}.  0FGL J1907.5+0602 was
found to pulse with a period of 106.6~ms, have a spin-down energy of
$\sim 2.8\times 10^{36}$ erg s$^{-1}$, and was given a preliminary
designation of PSR J1907+06. In this paper we derive a coherent timing
solution using 14 months of data which yields a more precise position
for the source, allowing detailed follow-up at other wavelengths,
including the detection of radio pulsations using the Arecibo 305-m
radio telescope.  Energy resolved light curves, the pulsed spectrum,
and off-pulse emission limits at the positions of the pulsar and PWN
centroid are presented.  We then report the detection of an X-ray
counterpart with the \textit{Chandra X-ray Observatory} and an upper
limit from \textit{ASCA}.  Finally, we discuss the pulsar's
relationship to the TeV source and to the potential birth sites SNR
G40.5$-$0.5 and the S74 H{\sc ii} region.

\section{Gamma-ray Pulsar Timing and Localization}
\label{sec:timing}

The discovery and initial pulse timing of PSR J1907+06 was reported by
\cite{BSP}.  The source position used in that analysis (R.A. =
286.965$^{\circ}$, Decl. = 6.022$^\circ$) was derived from an analysis
of the measured directions of LAT-detected photons in the on-pulse
phase interval from observations made from 2008 August 4 through
December 25. Here, we make use of a longer span of data and also apply
improved analysis methods to derive an improved timing ephemeris for
the pulsar as well as a more accurate source position.

For the timing and localization analysis, we selected ``diffuse''
class photons (events that passed the tightest background rejection
criteria \citep{LATinstrument}) with zenith angle $< 105^{\circ}$ as
is standard practice and chose the minimum energy and extraction
radius to optimize the significance of pulsations. We accepted photons
with $E>200$ MeV from within a radius of 0.7$^\circ$ of the nominal
source direction. We corrected these photon arrival times to
terrestrial time (TT) at the geocenter using the LAT Science Tool
\footnote{http://fermi.gsfc.nasa.gov/ssc/data/analysis/documentation/index.html}
\texttt{gtbary} in its geocenter mode.

\begin{figure}
\begin{center}
\includegraphics[scale=0.65,angle=-90]{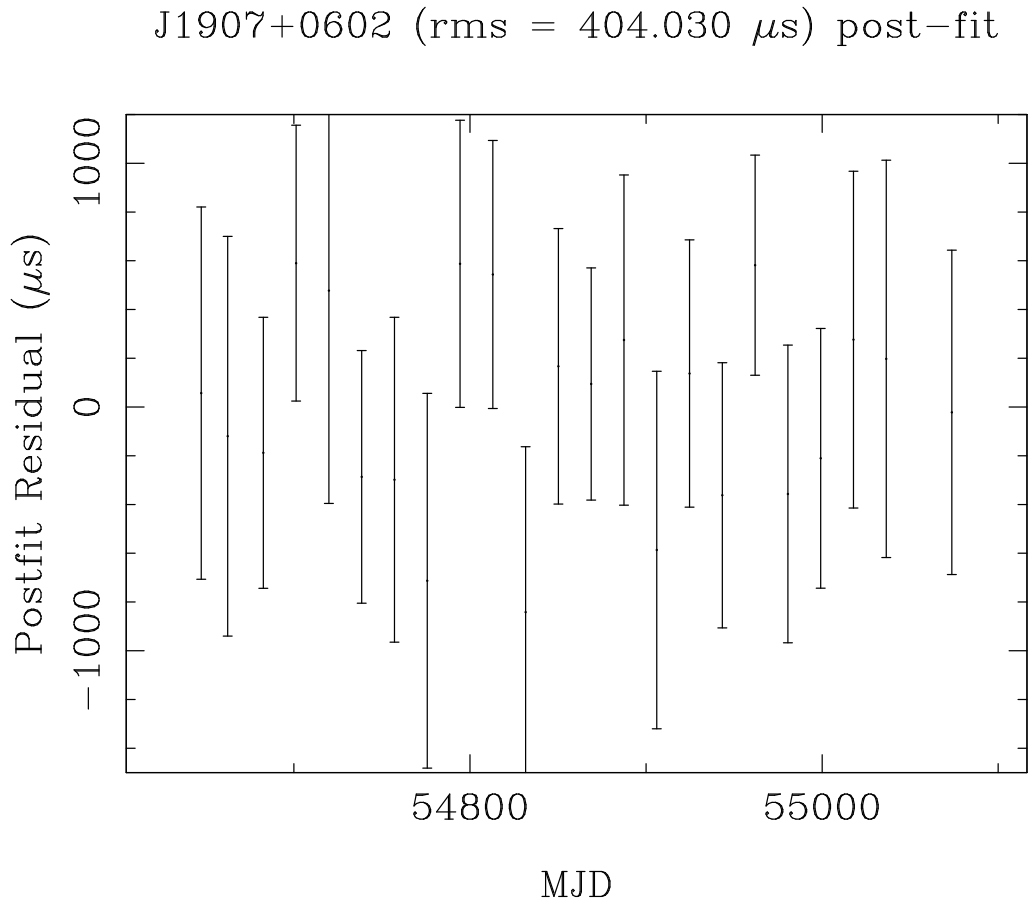}
\caption{Post-fit timing residuals for PSR J1907+0602. The reduced
chi-square of the fit is 0.5.}
\label{fig:timingresiduals}
\end{center}
\end{figure}

We fitted a timing model using \textsc{Tempo2}\citep{Hobbs2006} to 23
pulse times of arrival (TOAs) covering the interval 2008 June 30 to
2009 September 18. We note that during the on-orbit checkout period
(before 2008 August 4) several instrument configurations were tested
that affected the energy resolution and event reconstruction but had
no effect on the LAT timing. To determine the TOAs, we generated pulse
profiles by folding the photon times according to a provisional
ephemeris using polynomial coefficients generated by \textsc{Tempo2}
in its predictive mode (assuming a fictitious observatory at the
geocenter).  The TOAs were measured by cross correlating each pulse
profile with a kernel density template that was derived from fitting
the full mission dataset \citep{BlindTiming}. Finally, we fitted the
TOAs to a timing model that included position, frequency, and
frequency derivative. The resulting timing residuals are $0.4\, ms$
and are shown in Figure \ref{fig:timingresiduals}. The best-fit model
is displayed in Table \ref{tab:ephemeris}. The numbers in parentheses
are the errors in the last digit of the fitted parameters.  The errors
are statistical only, except for the position error, as described
below. The derived parameters of $\dot E$, $B$, and $\tau_c$ are
essentially unchanged with respect to those reported by \citet{BSP},
but the position has moved by 1.2\arcmin.

The statistical error on the position fit is $<1$\arcsec ;
however, this is an underestimate of the true error. For example, with
only one year of data, timing noise can perturb the position fit.  We
have performed a Monte Carlo analysis of these effects by simulating
fake residuals using the \textsc{fake} plugin for \textsc{Tempo2}.  We
generated models with a range of frequency second derivatives ($\pm
2\times 10^{-22} \mathrm{s}^{-3}$, the allowed magnitude for
$\ddot{\nu}$ in our fits) to simulate the effects of timing noise and
fitted them to timing models. Based on these simulations, we assigned
an additional systematic error on the position of 2\arcsec,
which we added in quadrature to the statistical error in Table
\ref{tab:ephemeris}.  As a result of the improved position estimate
provided by this timing analysis, we have adopted a more precise name
for the pulsar of PSR J1907+0602.

\section{Detection of Radio Pulsations}
 
To search for radio pulsations, we observed the timing position of PSR
J1907+0602 with the L-wide receiver on the Arecibo 305-m radio
telescope. On 2009 August 21 we made a 55-minute pointing with center
frequency 1.51 GHz and total bandwidth of 300 MHz, provided by three
Wideband Arecibo Pulsar Processors (WAPPs, \cite{Arecibo}), each
individually capable of processing 100 MHz. We divided this band into
512-channel spectra accumulated every 128 $\mu$s. The small positional
uncertainty of PSR J1907+0602 derived from the LAT timing means that a
single Arecibo pointing covers the whole region of interest.

After excising strong sources of radio-frequency interference with
\texttt{rfifind}, one of the routines of the PRESTO signal analysis
package \citep{Ransom2002}, we performed a search by folding the raw
data with the \textit{Fermi} timing model into 128-bin pulse
profiles. We then used the PRESTO routine \texttt{prepfold} to search
trial dispersion measures between 0 and 1000 pc cm$^{-3}$. We found a
pulsed signal with a signal-to-noise ratio $S/N=9.4$\footnote{This was
estimated using another software package, SIGPROC (a package developed
by Duncan Lorimer, see http://sigproc.sourceforge.net/), which
processes the bands separately and produced S/N of 4.2, 6.3 and 5.5
for the WAPPs centered at 1410, 1510 and 1610 MHz. Although S/N = 9.4
is close to the detection threshold for pulsars in a blind search, it
is much more significant in this case because of the reduced number of
trials in this search relative to a blind search.} and duty cycle of
about 0.03 at a dispersion measure $DM=82.1 \pm 1.1$ cm$^{-3}$
pc. This value was estimated by dividing the detection data into 3
sub-bands and making TOAs for each sub-band and fitting for the DM
with {\sc tempo}.

We applied the same technique for 4 different time segments of 12.5
minutes each and created a time of arrival for each of them. We then
estimated the barycentric periodicity of the detected signal from
these times of arrival.  This differs from the periodicity predicted
by the LAT ephemeris for the time of the observation by
$(-0.000005 \pm 0.000020)$ ms, i.e., the signals have the same
periodicity.

\begin{figure}
\begin{center}
\includegraphics[scale=0.5]{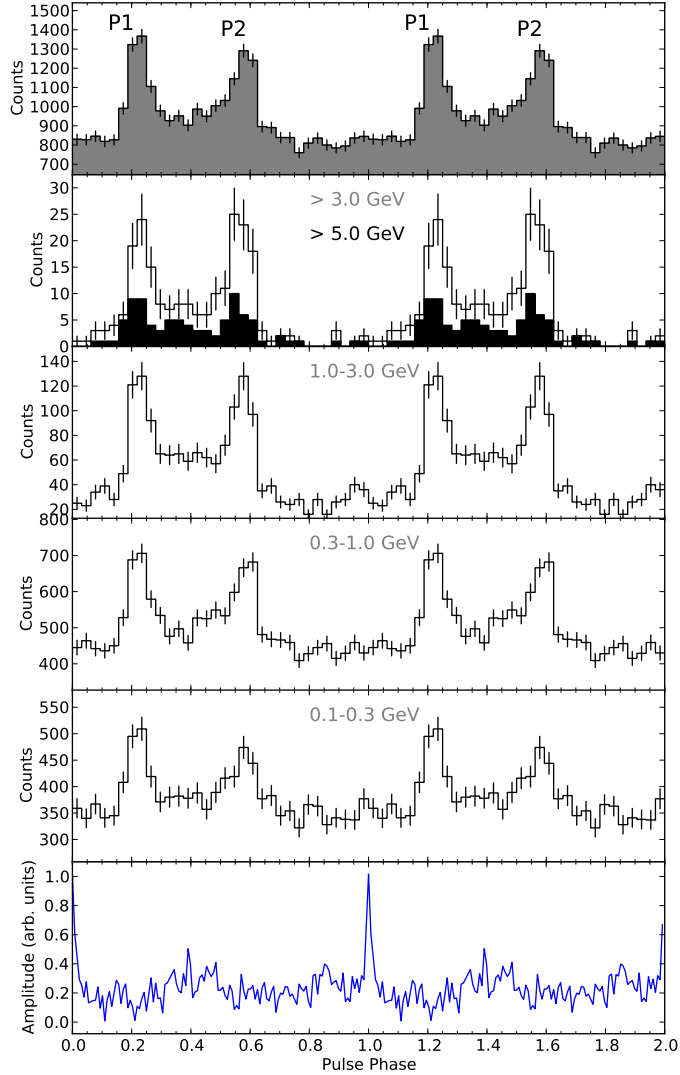}
\caption{Folded light curves of PSR J1907+0602 in 32 constant-width
bins for different energy bands and shown over two pulse periods with
the 1.4 GHz radio pulse profile plotted in the bottom panel. The top
panel of the figure shows the folded light curve for photons with $E >
0.1 $ GeV. The other panels show the pulse profiles in exclusive
energy ranges: $E > 3.0 $ GeV (with $E > 5.0$ GeV in black) in the
second panel from the top; 1.0 to 3.0 GeV in the next panel; 0.3 to
1.0 GeV in the fourth panel; and 0.1 to 0.3 GeV in the fifth panel. }
\label{fig:LC}
\end{center}
\end{figure}

Subsequent radio observations showed that the phase of the radio
pulses is exactly as predicted by the LAT ephemeris, apart from a
constant phase offset (depicted in Figure \ref{fig:LC})


A confirmation observation with twice the integration time (1.8 hr)
was made on 2009 September 4. The radio profile is shown in the bottom
panel of Figure. \ref{fig:LC} with an arbitrary intensity scale. The
pulsar is again detected with S/N of 3.4, 5.1, 7.3 and 8.6 at 1170,
1410, 1510 and 1610 MHz. The higher S/N at the higher frequencies
suggest a positive spectral index, similar to what has been observed
for PSR J1928+1746 \citep{Cordes2006}. However, this might instead be
due to scattering degrading the S/N at the lowest frequencies--- for
the band centered at 1610 MHz the pulse profile is distinctively
narrower (about 2\% at 50\% power) than at 1410 or 1510 MHz (about
3\%).  At 1170 MHz the profile is barely detectable but very broad.
This suggests an anomalously large scattering timescale for the DM of
the pulsar. Observations at higher frequencies will settle the issue
of the positive spectral index. For the 300 MHz centered at 1410 MHz,
where the detection is clear, we obtain a total S/N of 12.4.

With an antenna $T_\mathrm{sys} = 33$~K \citep[given by the
frequency-dependent antenna temperature of 25--27 K off the plane of
the Galaxy plus 6 K of Galactic emission in the specific direction of
the pulsar ][]{Haslam1982}, Gain = 10.5 K Jy$^{-1}$ and 2
polarizations, and an inefficiency factor of 12\% due to the 3-level
sampling of the WAPP correlators, we obtain for the first detection a
flux density at 1.4 GHz of $S_{1400}\,\simeq \,4.1 \,\mu$Jy and for
the second detection $S_{1400}\,\simeq \,3.1 \,\mu$Jy. These values
are consistent given the large relative uncertainties in the S/N
estimates and the varying effect of radio frequency interference; at
this DM, scintillation is not likely to cause a large variation in the
flux density.

The time-averaged flux density is $\simeq \,3.4 \,\mu$Jy. Using the
NE2001 model for the electron distribution in the Galaxy
\citep{Cordes2002}, we obtain from the pulsar's position and DM a
distance of 3.2 kpc with a nominal error of 20\%\citep{Cordes2002}.
The time-averaged flux density thus corresponds to a pseudo-luminosity
$L_{1400}\simeq 0.035$ mJy kpc$^2$.  This is fainter than the least
luminous young pulsar in the ATNF catalog (PSR J0205+6449, with a 1.4
GHz pseudo-luminosity of 0.5 mJy kpc$^2$). It is, however, more
luminous than the radio pulsations discovered through a deep search of
another pulsar first discovered by Fermi, PSR J1741$-$2054 which has
$L_{1400} \sim 0.025$ mJy kpc$^2$ \citep{Camilo2009}. These two
detections clearly demonstrate that some pulsars, as seen from the
Earth, can have extremely low apparent radio luminosities; i.e.,
similarly deep observations of other $\gamma$-ray selected pulsars
might detect additional very faint radio pulsars.  We note that these low
luminosities, which may well be the result of only a faint section of
the radio beam crossing the Earth, are much lower than what has often
been termed ``radio quiet'' in population synthesis models used to
estimate the ratio of ``radio-loud" to ``radio quiet" $\gamma$-ray
pulsars \citep[eg.][]{Gonthier2004}.

\section{Energy-Dependent Gamma-ray Pulse Profiles}

\label{sec:profiles}
The pulse profile and spectral results reported in this paper use the
survey data collected with the LAT from 2008 August 4 through 2009
September 18.  We selected ``diffuse'' class photons (see \S
\ref{sec:timing}) with energies $ E > 100$ MeV and, to limit
contamination from photons from Earth's limb, with zenith angle $<
105^{\circ}$.

To explore the dependence of the pulse profile on energy, we selected
an energy-dependent region of interest (ROI) with radius $\theta = 0.8
\times E^{-0.75} $ degrees, but constrained not to be outside the
range [$0.35^{\circ}, 1.5^{\circ}$]. We chose the upper bound to
minimize the contribution from nearby sources and Galactic diffuse
emission. The lower bound was selected in order to include more
photons from the wings of the point spread function (PSF) where the
extraction region is small enough to make the diffuse contribution
negligible.  Figure \ref{fig:LC} shows folded light curves of the
pulsar in 32 constant-width bins for different energy bands. We use
the centroid of the 1.4~GHz radio pulse profile to define phase
0.0. Two rotations are shown in each case. The top panel of the figure
shows the folded light curve for photons with $E > 0.1 $ GeV. The
$\gamma$-ray light curve shows two peaks, P1 at phase 0.220 $\pm$
0.002 which determines the offset with the radio peak, $\delta$. The
second peak in the $\gamma$-ray, P2, occurs at phase 0.580 $\pm$
0.003. The phase separation between the two peaks is $\Delta = 0.360
\pm 0.004$. The radio lead $\delta$ and gamma peak separation $\Delta$
values are in good agreement with the correlation predicted for outer
magnetosphere models, \citep{Romani1995} and observed for other young
pulsars (Figure 3 of \cite{Fermicatalog}).

Pulsed emission from the pulsar is clearly visible for energies $E >$
5 GeV with a chance probability of $\sim 4\times10^{-8}$. Pulsed
emission is detected for energies above 10 GeV with a confidence level
of 99.8\%. We have measured the integral and widths of the peaks as a
function of energy and have found no evidence for significant
evolution in shape or P1/P2 ratio with energy.  We note that the
pulsar is at low Galactic latitude ($b \sim -0.89^{\circ}$ ) where the
Galactic $\gamma$-ray diffuse emission is bright ( it has not been
subtracted from the light curves shown.)

\begin{figure}
\begin{center}
\includegraphics[scale=0.7]{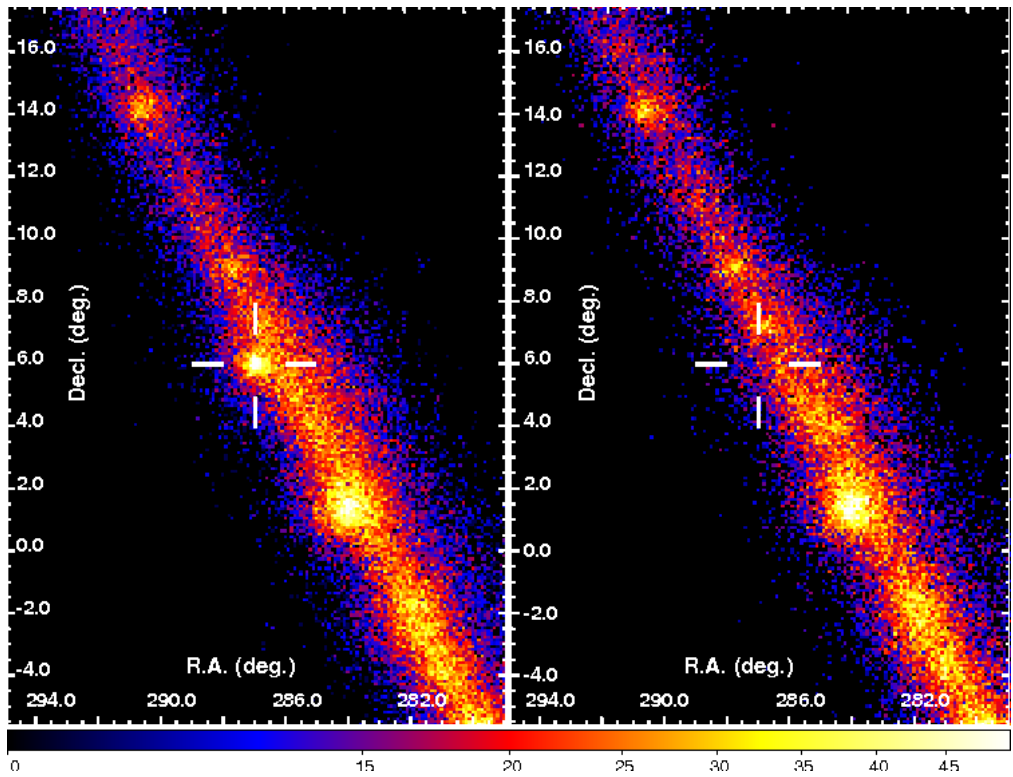}
\caption{The observed \textit{Fermi}-LAT counts map of the region
  around PSR J1907+0602. Left: ``on" pulse image, right: ``off'' pulse
  image. The open cross-hair marks the location of the pulsar. Color
  scale shows the counts per pixel.}
\label{fig:3images}
\end{center}
\end{figure}

Figure \ref{fig:3images} shows the observed LAT counts map of the
region around PSR J1907+0602.  We defined the ``on" pulse as pulse
phases 0.12 $\le \phi \le$ 0.68 and the ``off" pulse as its complement
(0.0 $ \le \phi <$ 0.12 and 0.68 $ < \phi \le$ 1.0).  We produced
on-pulse (left panel) and off-pulse (right panel) images, scaling the
off-pulse image by 1.27.  The figure indicates the complexity of the
region that must be treated in spectral fitting. Besides the pulsar
there are multiple point sources, Galactic, and extragalactic diffuse
contributions.

\section{Energy Spectrum}

\label{sec:spectrum}
The phase-averaged flux of the pulsar was obtained by performing a
maximum likelihood spectral analysis using the \textit{Fermi} LAT
science tool \texttt{gtlike}. Starting from the same data set
described in \S\ref{sec:profiles}, we selected photons from an ROI of
10 degrees around the pulsar position. Sources from a preliminary
version (based on 11 months of data) of the first \textit{Fermi} LAT
$\gamma-$ray catalog \citep{Fermicatalog} that are within 15 degree
ROI around the pulsar were modeled in this analysis. Spectra of
sources farther away than 5$^{\circ}$ from the pulsar were fixed at
the cataloged values. Sources within 5$^{\circ}$ degrees of the pulsar
were modeled with a simple power law. For each of the sources in the
5$^{\circ}$ degree region around the pulsar, we fixed the spectral
index at the value in the catalog and fitted for the
normalization. Two sources that are at a distance $> 5 ^{\circ}$
showed strong emission and were treated the same way as the sources
within 5$^{\circ}$. The Galactic diffuse emission (gll\_iemv02) and
the extragalactic diffuse background (isotropic\_iem\_v02) were
modeled as well\footnote{Descriptions of the models are available at
http://fermi.gsfc.nasa.gov/}.

The assumed spectral model for the pulsar is an exponentially cut-off
power law: $dN/dE = N_{o}\mbox{ } (E/E_{o})^{-\Gamma} \exp(-E/E_{c}).$
The resulting spectrum gives the total emission for the pulsar
assuming that the $\gamma$-ray emission is 100\% pulsed.  The unbinned
\texttt{gtlike} fit, using P6\_v3 instrument response functions
\citep{LATinstrument}, for the energy range $E \ge 100$ MeV gives a
phase-averaged spectrum of the following form:
\begin{equation}
\label{eqn:exppowerlaw}
  \frac{dN}{dE} = (7.06 \pm 0.43_{stat.} \, + (^{+0.004} _{-0.064})_{sys.}) \times 10^{-11}
  E^{-\Gamma}e^{-E/E_\mathrm{c}} \, \gamma \, \mathrm{cm}^{-2}\, \mathrm{s}^{-1}
  \,\mathrm{MeV}^{-1}
\end{equation}
where the photon index $\Gamma= 1.76 \pm 0.05_{stat.} + \, (^{+0.271}
_{-0.287})_{sys.}$ and the cutoff energy $E_{c} = 3.6 \pm 0.5_{stat.}
\, + (^{+0.72} _{-0.36})_{sys.}$ GeV.

The integrated energy flux from the pulsar in the energy range $E \ge
100$ MeV is $ F_{\gamma} = (3.12 \pm 0.15_{stat.} \, + (^{+0.16}
_{-0.15})_{sys.}) \times 10^{-10} \mathrm{ergs\;cm}^{-2}
\mathrm{s}^{-1}$. This yields a $\gamma$-ray luminosity of $
L_{\gamma} = 4\pi f_{\Omega}F_{\gamma}d^{2} = 3.8 \times 10^{35}
f_{\Omega}d_{3.2}^{2} \mathrm{ergs\;s}^{-1}$ above 100 MeV, where
$f_{\Omega}$ is an effective beaming factor and
$d^{}_{3.2}=d/(3.2){\rm kpc}$. This corresponds to an efficiency of
$\eta = L_{\gamma}/\dot{E} = 0.13 F_{\gamma} d_{3.2}^2$ for conversion
of spin-down power into $\gamma-$ray emission in this energy band.

\begin{figure}
\begin{center}
\includegraphics[scale=0.7]{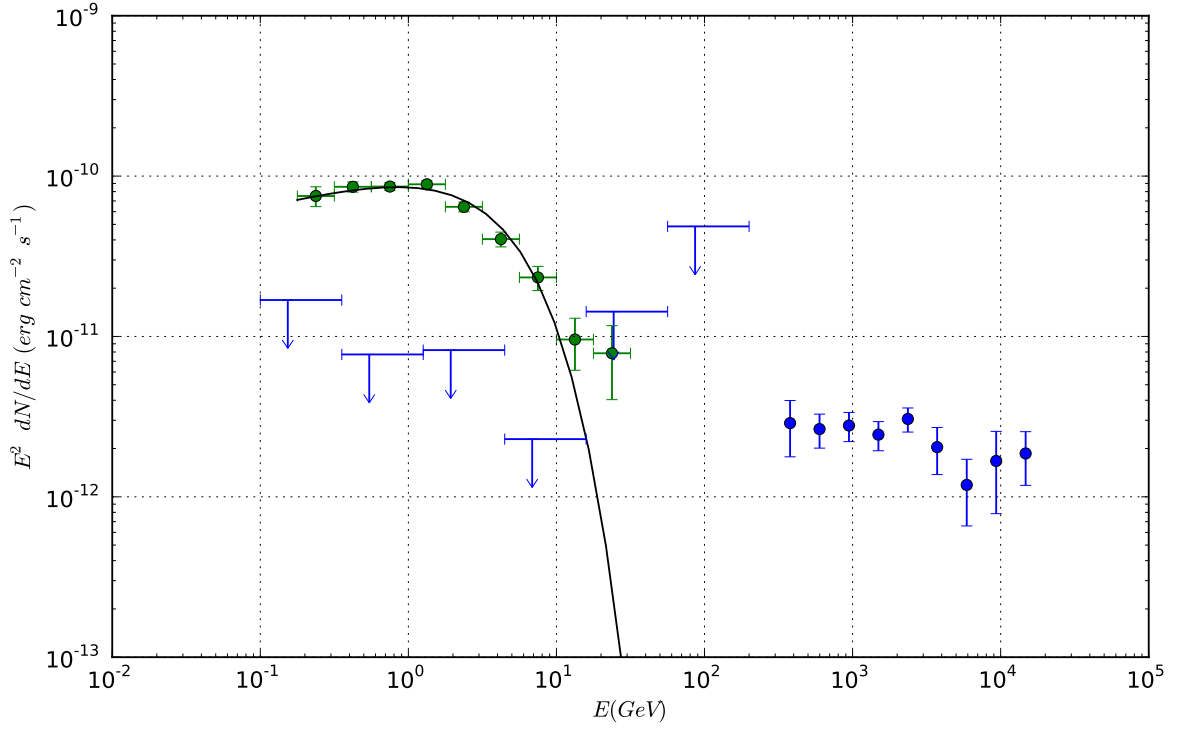}
\caption{Phase-averaged spectral energy distribution for PSR
  J1907+0602 (green circles). Blue circles are data from HESS for HESS
  J1908+063 TeV source. 2 $\sigma$ upper limits from Fermi for
  emission from this TeV source are shown in blue. The black line
  shows the spectral model for the pulsar (equation
  \ref{eqn:exppowerlaw}). The upper limits suggest that the spectrum
  of HESS J1908+063 has a low energy turnover between 20~GeV and
  300~GeV.}
\label{fig:spec}
\end{center}
\end{figure}

\begin{figure}
\begin{center}
\includegraphics[scale=0.8]{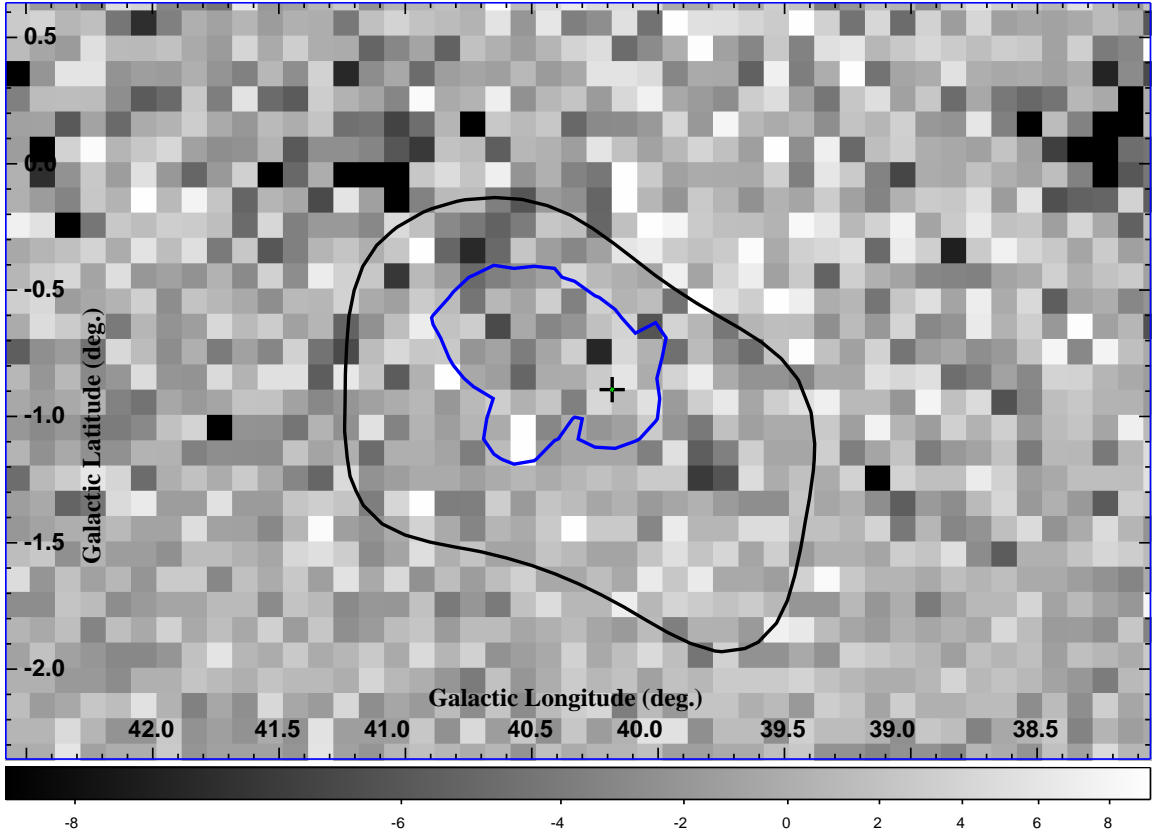}
\caption{Residual map of the region around PSR J1907+0602 in the
  off-pulse. The timing position of the pulsar is marked by the
  cross. The 5 $\sigma$ contours from Milagro (outer) and HESS (inner)
  are overlaid.}
\label{fig:residual-map}
\end{center}
\end{figure}

We set a 2$\sigma$ flux upper limit on $\gamma$-ray emission from the
pulsar in the off-pulse part of $F_{off} < 8.31 \times 10^{-8}\; {\rm
cm^{-2}}\; {\rm s^{-1}}$. In addition to the $\gamma$-ray spectrum
from the point-source pulsar PSR J1907+0602, we measured upper limits
on $\gamma$-ray flux from the extended source HESS J1908+063 in the
energy range 0.1--25 GeV. We performed binned likelihood analysis
using the \textit{Fermi} Science Tool \texttt{gtlike}. In this
analysis we assumed an extended source with gaussian width of
0.3$^{\circ}$ and $\gamma$-ray spectral index of $-2.1$ at the
location of the HESS source. The upper limits suggest that the
spectrum of HESS J1908+063 has a low energy turnover between 20~GeV
and 300~GeV. Figure \ref{fig:spec} shows the phase-averaged spectral
energy distribution for PSR J1907+0602 (green circles). On the same
figure we show data points from HESS for the TeV source HESS J1908+063
(blue circles) and the 2$\sigma$ upper limits from \textit{Fermi} for
emission from this TeV source. Figure \ref{fig:residual-map} shows an
off pulse residual map of the region around PSR J1907+0602. The timing
position of the pulsar is marked by the green cross. The 5$\sigma$
contours from Milagro (outer) and HESS (inner) are overlaid. As can be
seen from the residual map, there is no gamma-ray excess at the
location of either the pulsar or the PWN.

\section{X-ray Counterpart}

A 23~ks {\em ASCA} GIS exposure of the EGRET source GeV J1907+0557
revealed an $\sim 8^{\prime}\times 15^{\prime}$ region of possible
extended hard emission surrounding two point-like peaks lying $\sim
15^{\prime}$ to the southwest of PSR J1907+0602 \citep{RRK01} and no
other significant sources in the $44^{\prime}$ {\em ASCA} FOV. A 10~ks
{\em Chandra} ACIS-I image of the {\em ASCA} emission (ObsID 7049)
showed it to be dominated by a single hard point source, CXOU
J190718.6+054858 with no compact nebular structure and just a hint of
the several arcminute-scale emission seen by {\em ASCA}.  CXOU
J190718.6+054858 seemed to turn off for $\sim 2$ ks during the {\em
Chandra} exposure, suggesting that it may be a binary of some sort or
else a variable extragalactic source. There is no obvious optical
counterpart in the digital sky survey optical or 2MASS near infrared
images, nor in a I band image taken with the 2.4m Hiltner telescope at
MDM (Jules Halpern, private communication).  This strongly suggests
that it is not a nearby source. An absorbed power law fits the
spectrum of this source well, with absorption $n_H = 1.8^{+1.3}_{-0.9}
\times 10^{22} {\rm cm}^{-2}$ (90\% confidence region), a photon
spectral index $\Gamma = 0.9^{+0.6}_{-0.4}$, and an average 2--10 keV
flux of $4.4^{+0.7}_{-1.8}\times 10^{-13} {\rm erg}\,{\rm
cm}^{-2}\,{\rm s}^{-1}$ (68\% confidence region). The fit absorption
is similar to the estimated total Galactic absorption from the HEASARC
$nH$ tool of $1.6 \times 10^{22} {\rm cm}^{-2}$ based on the Dickey
and Lockman (1990) HI survey \citep{Dickey1990}, suggesting that an $n_H$ of $\sim
2\times 10^{22} {\rm cm}^{-2}$ is a reasonably conservative estimate
of interstellar absorption for sources deep in the plane along this
line of sight.

\begin{figure}
\begin{center}
\includegraphics[scale=0.9]{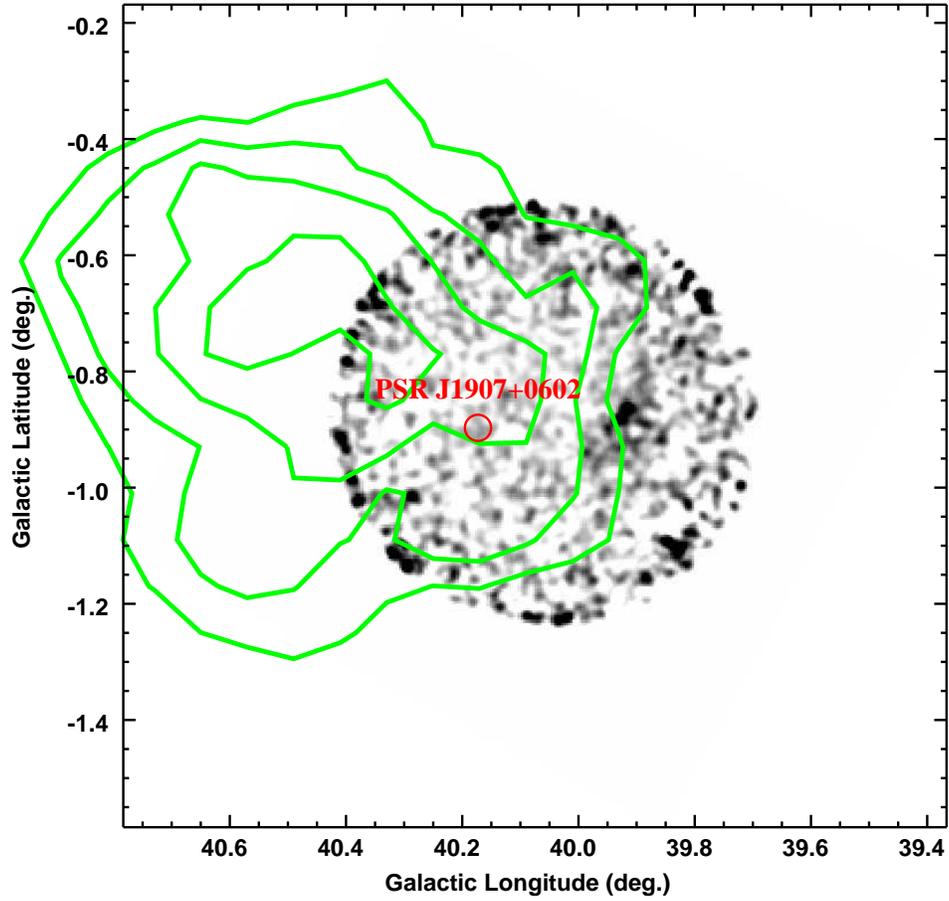}
\caption{{\em ASCA} GIS 2-10 keV image of the region around PSR
J1907+0602. The green contours are the 4-7 $\sigma$ significance
contours from HESS.}
\label{fig:asca}
\end{center}
\end{figure}

The timing position of LAT PSR J1907+0602 is in the central
$20^{\prime}$ of the {\em ASCA} GIS FOV (Figure \ref{fig:asca}). There
is no obvious emission in the {\em ASCA} image at the pulsar
position. Using the methodology of \cite{RRK01}, a 24 pixel radius
extraction region ($\sim 6^{\prime}$), and assuming an absorbed power
law spectrum with $n_H = 2\times 10^{22} {\rm cm}^{-2}$ and $\Gamma =
1.5$, we place a 90\% confidence upper limit on the 2--10 keV flux
$F_x < 5\times 10^{-14} {\rm erg}\,{\rm cm}^{-2}\,{\rm s}^{-1}$. This
suggests that for any reasonable absorption, the total unabsorbed
X-ray flux from the pulsar plus any arcminute-scale nebula is less
than $10^{-13} {\rm erg}\,{\rm cm}^{-2}\,{\rm s}^{-1}$.

\begin{figure}
\begin{center}
\includegraphics[scale=0.85]{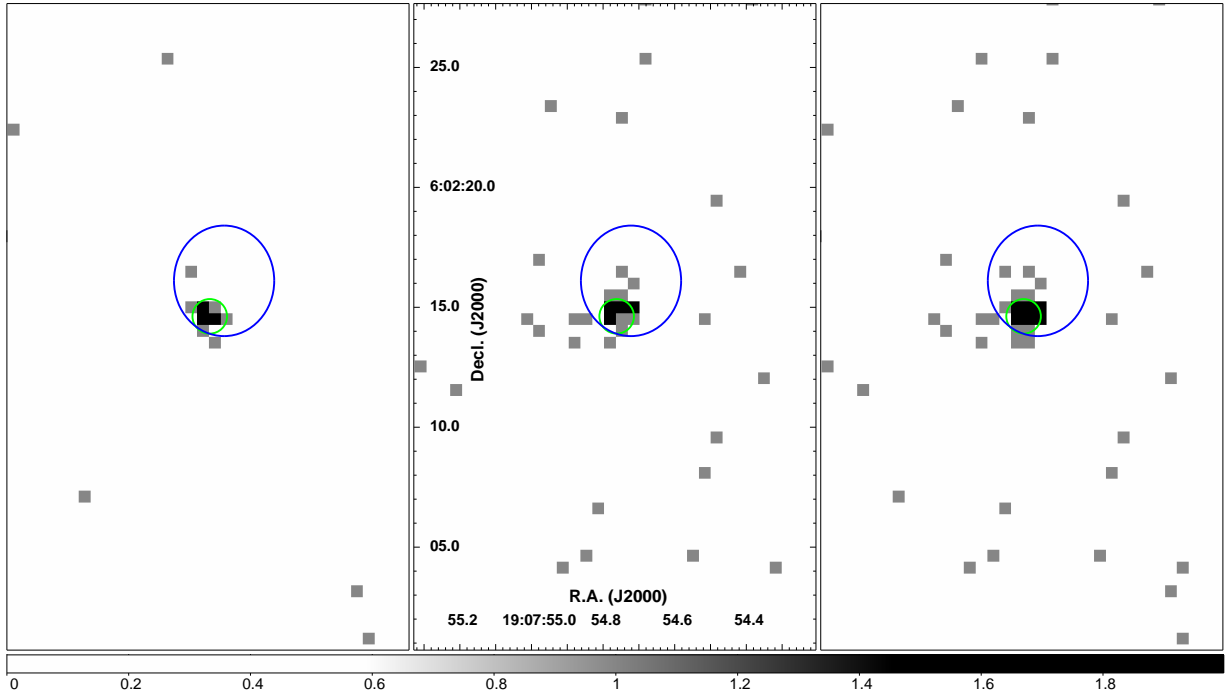}
\caption{{\em Chandra} ACIS images of PSR J1907+0602. The blue ellipse
  shows the uncertainty in the timing position. The green circle of
  radius 0.8\arcsec\ is twice the FWHM of the 5keV PSF at this
  position, and should contain roughly 80\% of the counts. The image
  at 0.75-2~keV (Left), 2-8~keV (Center) and 0.75-8~keV (right) is
  shown. Color scale shows the counts per pixel.}
\label{fig:chandra}
\end{center}
\end{figure}

\begin{figure}
\begin{center}
\includegraphics[scale=0.6,angle=-90]{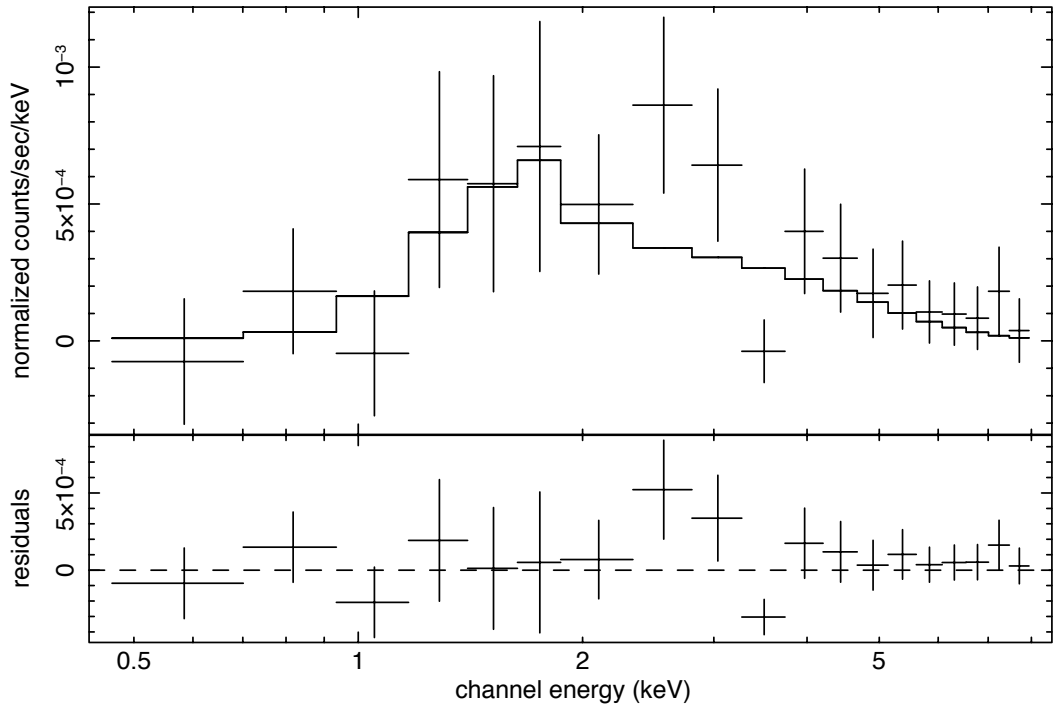}
\caption{{\em Chandra} X-ray spectrum of PSR J1907+0602.}
\label{fig:chandra_spec}
\end{center}
\end{figure}

PSR J1907+0602 was well outside of the FOV of the first {\em Chandra}
observation, and so we proposed for an observation centered on the
pulsar.  We obtained a 19~ks exposure with the ACIS-S detector (ObsID
11124). The time resolution of the ACIS-S detector on board {\em
Chandra} does not allow for pulse studies. The only source within an
arcminute of the timing position and the brightest source in the FOV
of the S3 chip is shown in Figure \ref{fig:chandra}. It is well within
errors of the timing position.  Examination of the X-ray image in
different energy bands showed virtually no detected flux below $\sim
1$keV and significant flux above 2.5 keV, suggesting a non-thermal
emission mechanism for much of the flux. A comparison of the spatial
distribution of counts between 0.75~keV and 2~keV to those between
2keV and 8keV shows some evidence for spatial extent beyond the point
spread function for the harder emission but not for the softer
emission. This would be consistent with an interpretation as
predominantly absorbed but thermal emission from a neutron star
surface surrounded by non-thermal emission from a compact pulsar wind
nebula, which is the typical situation for young pulsars
\citep[see][and references therein]{IsoNeutronStars}.  We plot the
{\em Chandra} 0.75-2keV, 2-8keV, and 0.75-8~keV images with an ellipse
showing the timing position uncertainty, and a circle with a radius of
$0.8^{\prime \prime}$. From a modeled PSF, we estimate 80\% of the
counts should be contained within this circle. While this seems to be
the case for the soft image, only roughly half the counts in the
harder image are contained within that radius. With only $\sim 12$
source counts in the 0.75-2~keV image within $6^{\prime \prime}$ and
$\sim 30$ source counts in the 2-8~keV image, quantitative statements
about source size and spectrum are difficult to make. We obtain a best
fit position for the nominal point source of R.A. =
19$^\mathrm{h}$07$^\mathrm{m}$54\fs76, decl. =
+06\degr02\arcmin14.6\arcsec\ and estimated error of 0.7\arcsec\ (an
additional 0.1\arcsec\ centroid fitting uncertainty added to the
nominal {\em Chandra} 0.6\arcsec\ uncertainty).  Using a $6^{\prime
\prime}$ radius extraction region and an annulus between 6\arcsec\ and
24\arcsec\ for background, we extracted source and background spectra
and fit them within XSPEC (Figure \ref{fig:chandra_spec}). A simple
power law plus absorption model fit the data well in the energy range
2-10 KeV, with best fit values $n_H = 1.3\times 10^{22} {\rm cm}^{-2}$
and $\Gamma = 1.6$, with a total flux of $2.3^{+0.6}_{-1.4}\times
10^{-14} {\rm erg}\, {\rm cm}^{-2} {\rm s}^{-1}$. The low count rates
and covariance between the absorption and photon index meant the
spectral parameters could not be simultaneously meaningfully
constrained. Fixing the spectral index $\Gamma=1.6$, a typical value
for compact pulsar wind nebulae \citep{IsoNeutronStars}, we obtain a
90\% confidence region for the absorption of $0.7-2.5 \times 10^{22}
{\rm cm}^{-2}$, consistent with a source a few kilo parsecs or more
away and with CXOU J190718.6+054858 discussed above.  We note that
with such an absorption a significant thermal component in the below 2
keV emission is neither required nor ruled out by the spectral
fitting.

\section{Discussion}

The dispersion measure from the radio detection suggests a distance of
3.2 kpc, with a nominal error of 20\%. However, there are many
outliers to the DM error distribution, although the largest fractional
errors tend to be from pulsars at high Galactic latitudes or very low
DMs \citep{Deller2009,Chatterjee2009}. For PSR J1907+0602, at a
latitude $b=-0.9^{\circ}$ with a moderate DM, the distance estimate is
likely to be reasonable. Since the apparent $\gamma$-ray pulsed
efficiency in the {\em Fermi} pass-band is well above the median for
other gamma-ray pulsars in \cite{pulsarcataog} (13\% compared to
7.5\%), it is worth checking secondary distance indicators to see if
the DM measure could be a significant overestimate of the true
distance.  We can use the X-ray observations of PSR J1907+0602 to do
this.  Several authors have noted a correlation between the X-ray
luminosity of young pulsars and their spin-down power (eg. Saito 1998,
Possenti et al. 2002, Li, Lu and Li 2007). Most of these have the
problem of using X-ray fluxes derived from the literature using a
variety of instruments with no uniform way of choosing spectral
extraction regions. This can be especially problematic with {\em
  Chandra} data, since faint, arcminute scale emission can easily be
overlooked. We compare our {\em ASCA} GIS upper limits to Figure 1 of
Cheng, Taam and Wang (2004) who used only {\em ASCA} GIS data to
derive their X-ray luminosity relationships. We see that the typical
X-ray luminosity in the {\em ASCA} band for a pulsar with $\dot E =
2\times 10^{36} {\rm erg}\, {\rm s}^{-1}$ is $L_x \sim 10^{33} -
10^{34} {\rm erg}\, {\rm s}^{-1}$ with all of the pulsars used in
their analysis with $\dot E > 10^{36} {\rm erg}\, {\rm s}^{-1}$ having
$L_x > 10^{32} {\rm erg}\, {\rm s}^{-1}$.  From these values and the
{\em ASCA} upper limit, we derive a lower limit for the distance to
LAT PSR J1907+0602 of $\sim 3$~kpc. From Figure 2 of Li, Lu and Li
(2007), who used {\em XMM-Newton} and {\em Chandra} derived values, we
see we can expect the luminosity to be between $\sim
10^{31.5}-10^{34.5} {\rm erg}\, {\rm s}^{-1}$.  From our detection
with {\em Chandra}, we again estimate a lower distance limit of $\sim
3$~kpc.  The ``best guess'' estimate from their relationship would
result in a distance of $\sim 13$kpc. We note that if we assume the
pulsed emission to be apparently isotropic (i.e. $f_{\Omega}=1$ as
simple outer gap models suggest should approximately be the case, see
\cite{Watters09}), a distance of 9~kpc would result in 100\%
$\gamma$-ray efficiency.

The derived timing position of PSR J1907+0602 is well inside the
extended HESS source, although $\sim 14^{\prime}$ southwest of the
centroid. The TeV source is therefore plausibly the wind nebula of PSR
J1907+0602. The physical size of this nebula is then $\simgt 40$ pc,
and the integrated luminosity above 1 TeV is $\simgt 40\%$ that of the
Crab, and in the MILAGRO band ($\sim 20$~TeV) at least twice that of
the Crab. There is a hint of some spatial dependence of the TeV
spectrum in the HESS data, with the harder emission ($> 2.5$~TeV)
peaking nearer the pulsar than the softer emission
\citep{HESS1908_2009}. If confirmed, this would be consistent with the
hardening of the TeV emission observed towards PSR B1823$-$13, thought
to be the pulsar powering HESS J1825$-$137
\citep{HESS_1825-137_2006}. This latter pulsar has a spin period,
characteristic age, and spin-down energy similar to PSR J1907+0602,
and is also located near the edge of its corresponding TeV nebula. We
also note that HESS J1825$-$137 subtends $\sim 1^{\circ}$ on the sky
and has a flux level above 1 TeV of around 20\% of the Crab. While the
overall spectrum of HESS J1825$-$137 is somewhat softer than the
spectrum of HESS J1908+063, near the pulsar its spectrum is similarly
hard.  At a distance of $\sim 4$~kpc, HESS J1825$-$137 has a
luminosity similar to the Crab TeV nebula, but with a much larger
physical size of $\sim 70$pc. Given the distance implied above and a
flux above 1 TeV $\sim 17$\% of the Crab, HESS J1908+063 is similar in
size and luminosity to HESS J1825$-$137.

At 20 TeV, HESS J1908+063 has a flux $\sim$80\% of the Crab, and so at
a distance $\simgt 1.5$ times that of the Crab, is much more luminous
at the highest energies. This is because there is no sign of a
high-energy cutoff or break, as is seen in many other TeV
nebulae. Aharonian et al. (2009) place a lower limit of 19.1 TeV on
any exponential cutoff to the spectrum. This implies that either the
spectrum is uncooled due to a very low nebular magnetic field ($\simlt
3 \mu$G, see, eg. \cite{Dejager2008}), an age much less than the
characteristic age of 19.5~kyr, or else there is a synchrotron cooling
break below the HESS band.

Our upper limits above a few GeV (Figure \ref{fig:spec}) requires
there to be a low energy turnover between 20~GeV and 300~GeV.  Given
the nominal PWN spectrum, we constrain the overall PWN flux to be $\le
\, 25\%$ of that of the pulsar. If only the HESS band is considered,
and assuming the DM distance, the TeV luminosity $L_\mathrm{PWN} =
5-8\% \dot E$.  However, since the TeV emission is generally thought
to come from a relic population of electrons the luminosity is likely
a function of the spin-down history of the pulsar rather than the
current spin-down luminosity \citep[eg.][]{Dejager2008}.  These
numbers support consistency of the association of the TeV source with
the pulsar, in the weak sense of not being discrepant with other
similar systems.

\subsection{On the possible association with SNR G40.5$-$0.5}

\begin{figure}
\begin{center}
\includegraphics[scale=0.6,angle=0]{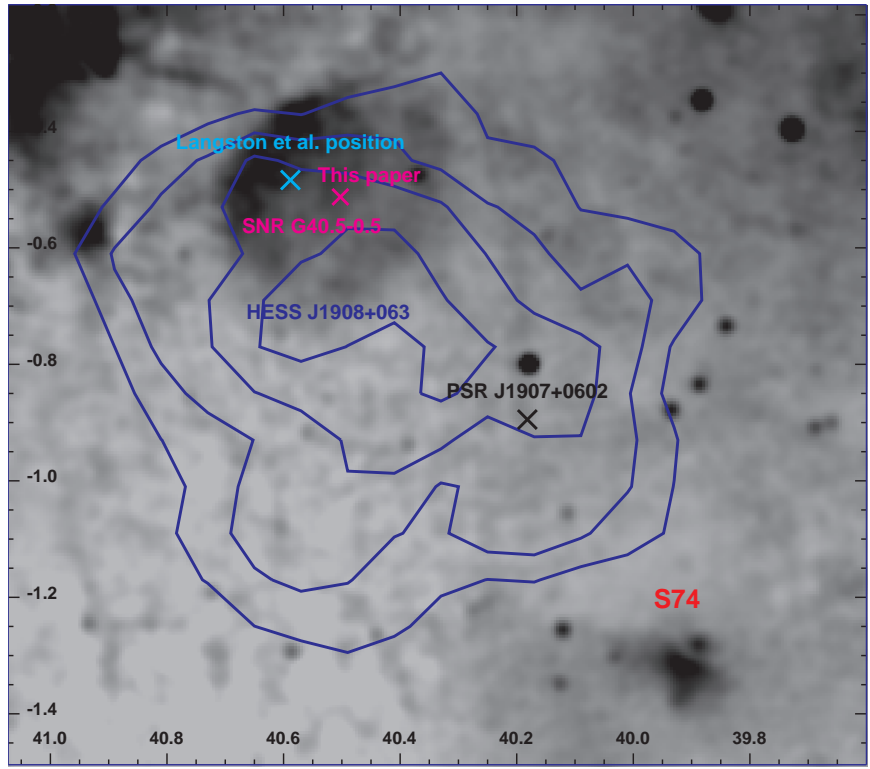}
\caption{VGPS 1420~MHz image of region in Galactic coordinates showing
relationship between SNR G40.5$-$0.5, HESS J1907+063 (blue contours
representing the 4,5,6 and 7$\sigma$ significance levels), the star
forming region S74, and PSR J1907+0602. }
\label{fig:vgps}
\end{center}
\end{figure}

The bulk of HESS J1908+063 is between PSR J1907+0602 and the young
radio SNR G40.5$-$0.5, suggesting a possible association. The distance
estimate \citep[$\sim3.4$~kpc][]{Yang2006} and age \citep{Downes1980}
estimates of SNR G40.5$-$0.5 are also consistent with those of PSR
J1907+0602.  If we use the usually assumed location for SNR G40.5-0.5
given by \cite{Langston2000} (RA=19$^\mathrm{h}$07$^\mathrm{m}11\fs9$,
Dec=6$^\circ35^{\prime}15^{\prime \prime}$), we get an angular
separation of $\sim 35^{\prime}$ between the timing position for the
pulsar and the SNR. However, this position for the SNR is from single
dish observations that were offset towards one bright side of the
nominal shell.  We use the VLA Galactic Plane Survey 1420~MHz image
\citep{Stil2006} of this region to estimate the SNR center to be
RA=19$^\mathrm{h}$07$^\mathrm{m}08\fs6$,
Dec=6$^\circ29^{\prime}53^{\prime \prime}$ (Figure \ref{fig:vgps})
which, for an assumed distance of 3.2~kpc, would give a separation of
$\sim$28 pc. Given the characteristic age of 19.5~kyr years, this
would require an average transverse velocity of $\sim 1400$ km/s.
While velocities about this high are seen in some cases (eg. PSR
B1508+55 has a transverse velocity of $\sim 1100$~km/s, Chatterjee et
al. 2005 \nocite{Chatterjee2005}), it is several times the average
pulsar velocity and many times higher than the local sound speed. We
note that pulsars with a braking index significantly less than $n=3$
assumed in the derivation of the characteristic age could have ages as
much as a factor of two greater \citep[see eg.][]{Kaspi2001}, and thus
a space velocity around half the above value may be all that is
required. But with any reasonable assumption of birthplace, distance,
and age, if the pulsar was born in SNR G40.5$-$0.5, any associated
X-ray or radio PWN should show a bow-shock and trail morphology, with
the trail likely pointing back towards the SNR center. Unfortunately,
the compactness and low number of counts in our {\em Chandra} image
precludes any definite statement about the PWN morphology. One arrives
at a different, and lower, minimum velocity if one assumes the pulsar
was born at the center of the TeV PWN and moved to its present
position, but the resulting velocity would still require a bow shock.

One can also get a pulsar offset towards the edge of a relic PWN if
there is a significant density gradient in the surrounding ISM. A
gradient will cause the supernova blast wave to propagate
asymmetrically. Where the density is higher, the reverse shock
propagating back to the explosion center will also be asymmetric. This
will tend to push the PWN away from the region of higher density
\citep{Blondin2001,Ferreira2008}. This has been invoked to explain the
offsets in the Vela X and HESS J1825$-$137 nebulae as well as several
others. Infrared and radio imaging of the region shows that HESS
J1908+063 borders on a shell of material surrounding the S74 HII
region, also known as the Lynds Bright Nebula 352.
\citet{Russeil2003} gives a kinematic distance of $3.0\pm 0.3$~kpc for
this star forming region, compatible with the pulsar distance. In this
scenario, the pulsar would not have to be highly supersonic to be at
the edge of a relic nebula, and would not have to be traveling away
from the center of the TeV emission.

A third, hybrid possibility is that SNR G40.5$-$0.5 is only a bright
segment of a much larger remnant, whose emission from the side near the
pulsar is confused with that from the molecular cloud. The asymmetry
would be explained by the difference in propagation speed in the lower
density ISM away from the molecular cloud.

Our current {\em Chandra} data are insufficient to distinguish between
the above scenarios. However, there is also the possibility of a
compact cometary radio nebula, such as is seen around PSR B1853+01 in
SNR W44 \citep{Frail1996} and PSR B0906$-$49 \citep{Gaensler1998}. In
addition, sensitive long wavelength radio imaging could reveal any
larger, faint SNR shells. Imaging with the EVLA and LOFAR of this
region is therefore highly desirable. The connection between the
pulsar and the TeV nebula could be further strengthened by a
confirmation of the spatio-spectral dependence of the nebula where the
spectrum hardens nearer to the pulsar.

\acknowledgments

The \textit{Fermi} LAT Collaboration acknowledges the generous support
of a number of agencies and institutes that have supported the {\em
Fermi} LAT Collaboration. These include the National Aeronautics and
Space Administration and the Department of Energy in the United
States, the Commissariat \`a l'Energie Atomique and the Centre
National de la Recherche Scientifique / Institut National de Physique
Nucl\'eaire et de Physique des Particules in France, the Agenzia
Spaziale Italiana and the Istituto Nazionale di Fisica Nucleare in
Italy, the Ministry of Education, Culture, Sports, Science and
Technology (MEXT), High Energy Accelerator Research Organization (KEK)
and Japan Aerospace Exploration Agency (JAXA) in Japan, and the
K. A. Wallenberg Foundation and the Swedish National Space Board in
Sweden.  The Arecibo Observatory is part of the National Astronomy and
Ionosphere Center, which is operated by Cornell University under a
cooperative agreement with the National Science Foundation. The
National Radio Astronomy Observatory is a facility of the National
Science Foundation Operated under cooperative agreement by Associated
Universities, Inc. Support for this work was provided by the National
Aeronautics and Space Administration through {\em Chandra} Award
Number GO6-7136X issued by the {\em Chandra} X-Ray Observatory Center,
which is operated by the Smithsonian Astrophysical Observatory for and
on behalf of NASA under contract NAS8-03060. This research has made
use of software provided by the {\em Chandra} X-Ray Center in the
application package CIAO. This research has made use of data obtained
from the High Energy Astrophysics Science Archive Research Center
(HEASARC), provided by NASA's Goddard Space Flight Center.

\bibliographystyle{apj}
\bibliography{Pulsar_Catalog_ALL_Refs}

\begin{center}
\begin{table}
\centering
\caption{Measured and Derived timing parameters of PSR J1907+0602}
\begin{tabular}{ll}
\hline\hline
\multicolumn{2}{c}{Fit and data-set} \\
\hline
Pulsar name\dotfill & J1907+0602 \\
MJD range\dotfill & 54647--55074 \\
Number of TOAs\dotfill & 23 \\
Rms timing residual ($\mu s$)\dotfill & 404 \\
\hline
\multicolumn{2}{c}{Measured Quantities} \\
\hline
Right ascension, $\alpha$\dotfill &  19:07:54.71(14) \\
Declination, $\delta$\dotfill & +06:02:16.1(23) \\
Pulse frequency, $\nu$ (s$^{-1}$)\dotfill & 9.3780713067(19) \\
First derivative of pulse frequency, $\dot{\nu}$ (s$^{-2}$)\dotfill & $-$7.6382(4)$\times 10^{-12}$ \\
Second derivative of pulse frequency, $\ddot{\nu}$ (s$^{-3}$)\dotfill & 2.5(6)$\times 10^{-22}$ \\
Epoch of frequency determination (MJD)\dotfill & 54800 \\
Dispersion measure, DM (cm$^{-3}$pc)\dotfill & 82.1(11)  \\
\hline
\multicolumn{2}{c}{Derived Quantities} \\
\hline
Characteristic age (kyr) \dotfill & 19.5 \\
Surface magnetic field strength (G) \dotfill & $3.1\times 10^{12}$ \\
$\dot{E}$ (erg s$^{-1}$) \dotfill & $2.8\times 10^{36}$ \\
\hline
\multicolumn{2}{c}{Assumptions} \\
\hline
Time units\dotfill & TDB \\
Solar system ephemeris model\dotfill & DE405 \\
\hline
\tablecomments{The numbers in parentheses are the errors in the last digit
of the fitted parameters.  The errors are statistical only, except for
the position error, as described in \S\ref{sec:timing}. The derived
parameters of $\dot{E}$, $B$, and $\tau_c$ are essentially unchanged
with respect to those reported by \citep{BSP}, but the position has
moved by 1.2 arcmin.}
\label{tab:ephemeris}
\end{tabular}
\end{table}
\end{center}

\end{document}

%% file: authors2.tex
\author{
A.~A.~Abdo\altaffilmark{2,3,1}, 
M.~Ackermann\altaffilmark{4}, 
M.~Ajello\altaffilmark{4}, 
L.~Baldini\altaffilmark{5}, 
J.~Ballet\altaffilmark{6}, 
G.~Barbiellini\altaffilmark{7,8}, 
D.~Bastieri\altaffilmark{9,10}, 
B.~M.~Baughman\altaffilmark{11}, 
K.~Bechtol\altaffilmark{4}, 
R.~Bellazzini\altaffilmark{5}
% ...
}
\altaffiltext{1}{Corresponding authors: A.~A.~Abdo, aous.abdo@nrl.navy.mil; M.~S.~E.~Roberts, malloryr@gmail.com; P.~M.~Saz~Parkinson, pablo@scipp.ucsc.edu; K.~S.~Wood, Kent.Wood@nrl.navy.mil.}
\altaffiltext{2}{Space Science Division, Naval Research Laboratory, Washington, DC 20375, USA}
\altaffiltext{3}{National Research Council Research Associate, National Academy of Sciences, Washington, DC 20001, USA}
\altaffiltext{4}{W. W. Hansen Experimental Physics Laboratory, Kavli Institute for Particle Astrophysics and Cosmology, Department of Physics and SLAC National Accelerator Laboratory, Stanford University, Stanford, CA 94305, USA}
\altaffiltext{5}{Istituto Nazionale di Fisica Nucleare, Sezione di Pisa, I-56127 Pisa, Italy}
\altaffiltext{6}{Laboratoire AIM, CEA-IRFU/CNRS/Universit\'e Paris Diderot, Service d'Astrophysique, CEA Saclay, 91191 Gif sur Yvette, France}
\altaffiltext{7}{Istituto Nazionale di Fisica Nucleare, Sezione di Trieste, I-34127 Trieste, Italy}
\altaffiltext{8}{Dipartimento di Fisica, Universit\`a di Trieste, I-34127 Trieste, Italy}
\altaffiltext{9}{Istituto Nazionale di Fisica Nucleare, Sezione di Padova, I-35131 Padova, Italy}
\altaffiltext{10}{Dipartimento di Fisica ``G. Galilei", Universit\`a di Padova, I-35131 Padova, Italy}
\altaffiltext{11}{Department of Physics, Center for Cosmology and Astro-Particle Physics, The Ohio State University, Columbus, OH 43210, USA}
% ...


%% file: LATPSRJ1907p0602_Revised.bbl
\begin{thebibliography}{39}
\expandafter\ifx\csname natexlab\endcsname\relax\def\natexlab#1{#1}\fi

\bibitem[{{Abdo} {et~al.}(2009{\natexlab{a}}){Abdo}, {Ackermann}, {Atwood},
  {Baldini}, {Ballet}, {Barbiellini}, {Baring}, {Bastieri}, \& {others}}]{BSP}
{Abdo}, A.~A. {et~al.} 2009{\natexlab{a}}, Science, 325, 840, (Blind Search
  Pulsars)

\bibitem[{{Abdo} {et~al.}(2009{\natexlab{b}}){Abdo}, {Ackermann}, {Atwood},
  {Baldini}, {Ballet}, {Barbiellini}, {Baring}, {Bastieri}, \& {others}}]{BSL}
---. 2009{\natexlab{b}}, \apjs, 183, 46, (Bright Source List)

\bibitem[{{Abdo} {et~al.}(2009{\natexlab{c}}){Abdo}, {Ackermann}, {Atwood},
  {Baldini}, {Ballet}, {Barbiellini}, {Baring}, {Bastieri}, \&
  {others}}]{Fermicatalog}
---. 2009{\natexlab{c}}, in prep (1st Fermi LAT Gamma-Ray Catalog)

\bibitem[{{Abdo} {et~al.}(2009{\natexlab{d}}){Abdo}, {Ackermann}, {Atwood},
  {Baldini}, {Ballet}, {Barbiellini}, {Baring}, {Bastieri}, \&
  {others}}]{pulsarcataog}
---. 2009{\natexlab{d}}, Submitted to ApJ (Fermi Catalog of Gamma-ray Pulsars)

\bibitem[{{Abdo} {et~al.}(2007){Abdo}, {Allen}, {Berley}, {Casanova}, {Chen},
  {Coyne}, {Dingus}, {Ellsworth}, {Fleysher}, {Fleysher}, {Gonzalez},
  {Goodman}, {Hays}, {Hoffman}, {Hopper}, {H{\"u}ntemeyer}, {Kolterman},
  {Lansdell}, {Linnemann}, {McEnery}, {Mincer}, {Nemethy}, {Noyes}, {Ryan},
  {Saz Parkinson}, {Shoup}, {Sinnis}, {Smith}, {Sullivan}, {Vasileiou},
  {Walker}, {Williams}, {Xu}, \& {Yodh}}]{Milagro07}
---. 2007, \apjl, 664, L91, (The Milagro Collaboration)

\bibitem[{{Aharonian} {et~al.}(2009){Aharonian}, {Akhperjanian}, {Anton},
  {Barres de Almeida}, {Bazer-Bachi}, {Becherini}, {Behera}, {Benbow},
  {Bernl{\"o}hr}, {Boisson}, {Bochow}, {Borrel}, {Braun}, {Brion}, {Brucker},
  {Brun}, {B{\"u}hler}, {Bulik}, {B{\"u}sching}, {Boutelier}, {Carrigan},
  {Chadwick}, {Charbonnier}, {Chaves}, {Cheesebrough}, {Chounet}, {Clapson},
  {Coignet}, {Dalton}, {Daniel}, {Degrange}, {Deil}, {Dickinson},
  {Djannati-Ata{\"i}}, {Domainko}, {O'C.~Drury}, {Dubois}, {Dubus}, {Dyks},
  {Dyrda}, {Egberts}, {Emmanoulopoulos}, {Espigat}, {Farnier}, {Feinstein},
  {Fiasson}, {F{\"o}rster}, {Fontaine}, {F{\"u}{\ss}ling}, {Gabici}, {Gallant},
  {G{\'e}rard}, {Giebels}, {Glicenstein}, {Gl{\"u}ck}, {Goret}, {Hauser},
  {Hauser}, {Heinz}, {Heinzelmann}, {Henri}, {Hermann}, {Hinton}, {Hoffmann},
  {Hofmann}, {Holleran}, {Hoppe}, {Horns}, {Jacholkowska}, {de Jager}, {Jung},
  {Katarzy{\'n}ski}, {Katz}, {Kaufmann}, {Kendziorra}, {Kerschhaggl},
  {Khangulyan}, {Kh{\'e}lifi}, {Keogh}, {Komin}, {Kosack}, {Lamanna}, {Lenain},
  {Lohse}, {Marandon}, {Martin}, {Martineau-Huynh}, {Marcowith}, {Maurin},
  {McComb}, {Medina}, {Moderski}, {Moulin}, {Naumann-Godo}, {de Naurois},
  {Nedbal}, {Nekrassov}, {Niemiec}, {Nolan}, {Ohm}, {Olive}, {de O{\~n}a
  Wilhelmi}, {Orford}, {Ostrowski}, {Panter}, {Paz Arribas}, {Pedaletti},
  {Pelletier}, {Petrucci}, {Pita}, {P{\"u}hlhofer}, {Punch}, {Quirrenbach},
  {Raubenheimer}, {Raue}, {Rayner}, {Renaud}, {Reimer}, {Rieger}, {Ripken},
  {Rob}, {Rosier-Lees}, {Rowell}, {Rudak}, {Rulten}, {Ruppel}, {Sahakian},
  {Santangelo}, {Schlickeiser}, {Sch{\"o}ck}, {Schr{\"o}der}, {Schwanke},
  {Schwarzburg}, {Schwemmer}, {Shalchi}, {Skilton}, {Sol}, {Spangler},
  {Stawarz}, {Steenkamp}, {Stegmann}, {Superina}, {Tam}, {Tavernet}, {Terrier},
  {Tibolla}, {van Eldik}, {Vasileiadis}, {Venter}, {Venter}, {Vialle},
  {Vincent}, {Vivier}, {V{\"o}lk}, {Volpe}, {Wagner}, {Ward}, {Zdziarski}, \&
  {Zech}}]{HESS1908_2009}
{Aharonian}, F. {et~al.} 2009, \aap, 499, 723

\bibitem[{{Aharonian} {et~al.}(2006){Aharonian}, {Akhperjanian}, {Bazer-Bachi},
  {Beilicke}, {Benbow}, {Berge}, {Bernl{\"o}hr}, {Boisson}, {Bolz}, {Borrel},
  {Braun}, {Brown}, {B{\"u}hler}, {B{\"u}sching}, {Carrigan}, {Chadwick},
  {Chounet}, {Cornils}, {Costamante}, {Degrange}, {Dickinson},
  {Djannati-Ata{\"i}}, {O'C.~Drury}, {Dubus}, {Egberts}, {Emmanoulopoulos},
  {Espigat}, {Feinstein}, {Ferrero}, {Fiasson}, {Fontaine}, {Funk}, {Funk},
  {F{\"u}{\ss}ling}, {Gallant}, {Giebels}, {Glicenstein}, {Goret},
  {Hadjichristidis}, {Hauser}, {Hauser}, {Heinzelmann}, {Henri}, {Hermann},
  {Hinton}, {Hoffmann}, {Hofmann}, {Holleran}, {Horns}, {Jacholkowska}, {de
  Jager}, {Kendziorra}, {Kh{\'e}lifi}, {Komin}, {Konopelko}, {Kosack},
  {Latham}, {Le Gallou}, {Lemi{\`e}re}, {Lemoine-Goumard}, {Lohse}, {Martin},
  {Martineau-Huynh}, {Marcowith}, {Masterson}, {Maurin}, {McComb}, {Moulin},
  {de Naurois}, {Nedbal}, {Nolan}, {Noutsos}, {Orford}, {Osborne}, {Ouchrif},
  {Panter}, {Pelletier}, {Pita}, {P{\"u}hlhofer}, {Punch}, {Raubenheimer},
  {Raue}, {Rayner}, {Reimer}, {Reimer}, {Ripken}, {Rob}, {Rolland}, {Rowell},
  {Sahakian}, {Santangelo}, {Saug{\'e}}, {Schlenker}, {Schlickeiser},
  {Schr{\"o}der}, {Schwanke}, {Schwarzburg}, {Shalchi}, {Sol}, {Spangler},
  {Spanier}, {Steenkamp}, {Stegmann}, {Superina}, {Tavernet}, {Terrier},
  {Th{\'e}oret}, {Tluczykont}, {van Eldik}, {Vasileiadis}, {Venter}, {Vincent},
  {V{\"o}lk}, {Wagner}, \& {Ward}}]{HESS_1825-137_2006}
---. 2006, \aap, 460, 365

\bibitem[{{Atwood} {et~al.}(2009){Atwood}, {Abdo}, {Ackermann}, {Althouse},
  {Anderson}, {Axelsson}, {Baldini}, {Ballet}, {Band}, {Barbiellini},
  {Bartelt}, {Bastieri}, {Baughman}, {Bechtol}, {B{\'e}d{\'e}r{\`e}de},
  {Bellardi}, {Bellazzini}, {Berenji}, {Bignami}, {Bisello}, {Bissaldi},
  {Blandford}, {Bloom}, {Bogart}, {Bonamente}, {Bonnell}, {Borgland},
  {Bouvier}, {Bregeon}, {Brez}, {Brigida}, {Bruel}, {Burnett}, {Busetto},
  {Caliandro}, {Cameron}, {Caraveo}, {Carius}, {Carlson}, {Casandjian},
  {Cavazzuti}, {Ceccanti}, {Cecchi}, {Charles}, {Chekhtman}, {Cheung},
  {Chiang}, {Chipaux}, {Cillis}, {Ciprini}, {Claus}, {Cohen-Tanugi},
  {Condamoor}, {Conrad}, {Corbet}, {Corucci}, {Costamante}, {Cutini}, {Davis},
  {Decotigny}, {DeKlotz}, {Dermer}, {de Angelis}, {Digel}, {do Couto e Silva},
  {Drell}, {Dubois}, {Dumora}, {Edmonds}, {Fabiani}, {Farnier}, {Favuzzi},
  {Flath}, {Fleury}, {Focke}, {Funk}, {Fusco}, {Gargano}, {Gasparrini},
  {Gehrels}, {Gentit}, {Germani}, {Giebels}, {Giglietto}, {Giommi}, {Giordano},
  {Glanzman}, {Godfrey}, {Grenier}, {Grondin}, {Grove}, {Guillemot}, {Guiriec},
  {Haller}, {Harding}, {Hart}, {Hays}, {Healey}, {Hirayama}, {Hjalmarsdotter},
  {Horn}, {Hughes}, {J{\'o}hannesson}, {Johansson}, {Johnson}, {Johnson},
  {Johnson}, {Johnson}, {Kamae}, {Katagiri}, {Kataoka}, {Kavelaars}, {Kawai},
  {Kelly}, {Kerr}, {Klamra}, {Kn{\"o}dlseder}, {Kocian}, {Komin}, {Kuehn},
  {Kuss}, {Landriu}, {Latronico}, {Lee}, {Lee}, {Lemoine-Goumard}, {Lionetto},
  {Longo}, {Loparco}, {Lott}, {Lovellette}, {Lubrano}, {Madejski}, {Makeev},
  {Marangelli}, {Massai}, {Mazziotta}, {McEnery}, {Menon}, {Meurer},
  {Michelson}, {Minuti}, {Mirizzi}, {Mitthumsiri}, {Mizuno}, {Moiseev},
  {Monte}, {Monzani}, {Moretti}, {Morselli}, {Moskalenko}, {Murgia},
  {Nakamori}, {Nishino}, {Nolan}, {Norris}, {Nuss}, {Ohno}, {Ohsugi}, {Omodei},
  {Orlando}, {Ormes}, {Paccagnella}, {Paneque}, {Panetta}, {Parent}, {Pearce},
  {Pepe}, {Perazzo}, {Pesce-Rollins}, {Picozza}, {Pieri}, {Pinchera}, {Piron},
  {Porter}, {Poupard}, {Rain{\`o}}, {Rando}, {Rapposelli}, {Razzano}, {Reimer},
  {Reimer}, {Reposeur}, {Reyes}, {Ritz}, {Rochester}, {Rodriguez}, {Romani},
  {Roth}, {Russell}, {Ryde}, {Sabatini}, {Sadrozinski}, {Sanchez}, {Sander},
  {Sapozhnikov}, {Parkinson}, {Scargle}, {Schalk}, {Scolieri}, {Sgr{\`o}},
  {Share}, {Shaw}, {Shimokawabe}, {Shrader}, {Sierpowska-Bartosik}, {Siskind},
  {Smith}, {Smith}, {Spandre}, {Spinelli}, {Starck}, {Stephens}, {Strickman},
  {Strong}, {Suson}, {Tajima}, {Takahashi}, {Takahashi}, {Tanaka}, {Tenze},
  {Tether}, {Thayer}, {Thayer}, {Thompson}, {Tibaldo}, {Tibolla}, {Torres},
  {Tosti}, {Tramacere}, {Turri}, {Usher}, {Vilchez}, {Vitale}, {Wang},
  {Watters}, {Winer}, {Wood}, {Ylinen}, \& {Ziegler}}]{LATinstrument}
{Atwood}, W.~B. {et~al.} 2009, \apj, 697, 1071, (LAT)

\bibitem[{{Blondin} {et~al.}(2001){Blondin}, {Chevalier}, \&
  {Frierson}}]{Blondin2001}
{Blondin}, J.~M., {Chevalier}, R.~A., \& {Frierson}, D.~M. 2001, \apj, 563, 806

\bibitem[{{Camilo} {et~al.}(2009){Camilo}, {Ray}, {Ransom}, {Burgay},
  {Johnson}, {Kerr}, {Gotthelf}, {Halpern}, {Reynolds}, {Romani}, {Demorest},
  {Johnston}, {van Straten}, {Saz Parkinson}, {Ziegler}, {Dormody}, {Thompson},
  {Smith}, {Harding}, {Abdo}, {Crawford}, {Freire}, {Keith}, {Kramer},
  {Roberts}, {Weltevrede}, \& {Wood}}]{Camilo2009}
{Camilo}, F. {et~al.} 2009, \apj, in press (arXiv:0908.2626) - Radio detection
  of two gamma-ray pulsars

\bibitem[{{Chatterjee} {et~al.}(2009){Chatterjee}, {Brisken}, {Vlemmings},
  {Goss}, {Lazio}, {Cordes}, {Thorsett}, {Fomalont}, {Lyne}, \&
  {Kramer}}]{Chatterjee2009}
{Chatterjee}, S. {et~al.} 2009, \apj, 698, 250

\bibitem[{{Chatterjee} {et~al.}(2005){Chatterjee}, {Vlemmings}, {Brisken},
  {Lazio}, {Cordes}, {Goss}, {Thorsett}, {Fomalont}, {Lyne}, \&
  {Kramer}}]{Chatterjee2005}
---. 2005, \apjl, 630, L61

\bibitem[{{Cordes} {et~al.}(2006){Cordes}, {Freire}, {Lorimer}, {Camilo},
  {Champion}, {Nice}, {Ramachandran}, {Hessels}, {Vlemmings}, {van Leeuwen},
  {Ransom}, {Bhat}, {Arzoumanian}, {McLaughlin}, {Kaspi}, {Kasian}, {Deneva},
  {Reid}, {Chatterjee}, {Han}, {Backer}, {Stairs}, {Deshpande}, \&
  {Faucher-Gigu{\`e}re}}]{Cordes2006}
{Cordes}, J.~M. {et~al.} 2006, \apj, 637, 446

\bibitem[{{Cordes} \& {Lazio}(2002)}]{Cordes2002}
{Cordes}, J.~M., \& {Lazio}, T.~J.~W. 2002, ArXiv e-prints,
  (arXiv:astro-ph/0207156)

\bibitem[{{de Jager}(2008)}]{Dejager2008}
{de Jager}, O.~C. 2008, \apjl, 678, L113

\bibitem[{{Deller} {et~al.}(2009){Deller}, {Tingay}, {Bailes}, \&
  {Reynolds}}]{Deller2009}
{Deller}, A.~T., {Tingay}, S.~J., {Bailes}, M., \& {Reynolds}, J.~E. 2009,
  \apj, 701, 1243

\bibitem[{{Dickey} \& {Lockman}(1990)}]{Dickey1990}
{Dickey}, J.~M., \& {Lockman}, F.~J. 1990, \araa, 28, 215

\bibitem[{{Dowd} {et~al.}(2000){Dowd}, {Sisk}, \& {Hagen}}]{Arecibo}
{Dowd}, A., {Sisk}, W., \& {Hagen}, J. 2000, in Astronomical Society of the
  Pacific Conference Series, Vol. 202, IAU Colloq. 177: Pulsar Astronomy - 2000
  and Beyond, ed. M.~{Kramer}, N.~{Wex}, \& R.~{Wielebinski}, 275

\bibitem[{{Downes} {et~al.}(1980){Downes}, {Salter}, \& {Pauls}}]{Downes1980}
{Downes}, A.~J.~B., {Salter}, C.~J., \& {Pauls}, T. 1980, \aap, 92, 47

\bibitem[{{Ferreira} \& {de Jager}(2008)}]{Ferreira2008}
{Ferreira}, S.~E.~S., \& {de Jager}, O.~C. 2008, \aap, 478, 17

\bibitem[{{Frail} {et~al.}(1996){Frail}, {Giacani}, {Goss}, \&
  {Dubner}}]{Frail1996}
{Frail}, D.~A., {Giacani}, E.~B., {Goss}, W.~M., \& {Dubner}, G. 1996, \apjl,
  464, L165+

\bibitem[{{Gaensler} {et~al.}(1998){Gaensler}, {Stappers}, {Frail}, \&
  {Johnston}}]{Gaensler1998}
{Gaensler}, B.~M., {Stappers}, B.~W., {Frail}, D.~A., \& {Johnston}, S. 1998,
  \apjl, 499, L69+

\bibitem[{{Gonthier} {et~al.}(2004){Gonthier}, {Van Guilder}, \&
  {Harding}}]{Gonthier2004}
{Gonthier}, P.~L., {Van Guilder}, R., \& {Harding}, A.~K. 2004, \apj, 604, 775

\bibitem[{{Hartman} {et~al.}(1999){Hartman}, {Bertsch}, {Bloom}, {Chen},
  {Deines-Jones}, {Esposito}, {Fichtel}, {Friedlander}, {Hunter}, {McDonald},
  {Sreekumar}, {Thompson}, {Jones}, {Lin}, {Michelson}, {Nolan}, {Tompkins},
  {Kanbach}, {Mayer-Hasselwander}, {M{\"u}cke}, {Pohl}, {Reimer}, {Kniffen},
  {Schneid}, {von Montigny}, {Mukherjee}, \& {Dingus}}]{3rdCat}
{Hartman}, R.~C. {et~al.} 1999, \apjs, 123, 79

\bibitem[{{Haslam} {et~al.}(1982){Haslam}, {Salter}, {Stoffel}, \&
  {Wilson}}]{Haslam1982}
{Haslam}, C.~G.~T., {Salter}, C.~J., {Stoffel}, H., \& {Wilson}, W.~E. 1982,
  \aaps, 47, 1

\bibitem[{{Hobbs} {et~al.}(2006){Hobbs}, {Edwards}, \&
  {Manchester}}]{Hobbs2006}
{Hobbs}, G., {Edwards}, R., \& {Manchester}, R. 2006, Chinese Journal of
  Astronomy and Astrophysics Supplement, 6, 189

\bibitem[{{Kaspi} {et~al.}(2001){Kaspi}, {Roberts}, {Vasisht}, {Gotthelf},
  {Pivovaroff}, \& {Kawai}}]{Kaspi2001}
{Kaspi}, V.~M., {Roberts}, M.~E., {Vasisht}, G., {Gotthelf}, E.~V.,
  {Pivovaroff}, M., \& {Kawai}, N. 2001, \apj, 560, 371

\bibitem[{{Kaspi} {et~al.}(2006){Kaspi}, {Roberts}, \&
  {Harding}}]{IsoNeutronStars}
{Kaspi}, V.~M., {Roberts}, M.~S.~E., \& {Harding}, A.~K. 2006, in Compact
  stellar X-ray sources, ed. W.~H.~G. {Lewin} \& M.~{van der Klis}, 279--339

\bibitem[{{Lamb} \& {Macomb}(1997)}]{Lamb1997}
{Lamb}, R.~C., \& {Macomb}, D.~J. 1997, \apj, 488, 872

\bibitem[{{Langston} {et~al.}(2000){Langston}, {Minter}, {D'Addario},
  {Eberhardt}, {Koski}, \& {Zuber}}]{Langston2000}
{Langston}, G., {Minter}, A., {D'Addario}, L., {Eberhardt}, K., {Koski}, K., \&
  {Zuber}, J. 2000, \aj, 119, 2801

\bibitem[{{Ransom} {et~al.}(2002){Ransom}, {Eikenberry}, \&
  {Middleditch}}]{Ransom2002}
{Ransom}, S.~M., {Eikenberry}, S.~S., \& {Middleditch}, J. 2002, \aj, 124, 1788

\bibitem[{{Ray} {et~al.}(2009)}]{BlindTiming}
{Ray}, P.~S., {et~al.} 2009, \apj, in prep (Precise Timing of Fermi Gamma-Ray
  Pulsars)

\bibitem[{{Roberts} {et~al.}(2001){Roberts}, {Romani}, \& {Kawai}}]{RRK01}
{Roberts}, M.~S.~E., {Romani}, R.~W., \& {Kawai}, N. 2001, \apjs, 133, 451

\bibitem[{{Romani} \& {Yadigaroglu}(1995)}]{Romani1995}
{Romani}, R.~W., \& {Yadigaroglu}, I.-A. 1995, \apj, 438, 314

\bibitem[{{Russeil}(2003)}]{Russeil2003}
{Russeil}, D. 2003, \aap, 397, 133

\bibitem[{{Stil} {et~al.}(2006){Stil}, {Taylor}, {Dickey}, {Kavars}, {Martin},
  {Rothwell}, {Boothroyd}, {Lockman}, \& {McClure-Griffiths}}]{Stil2006}
{Stil}, J.~M. {et~al.} 2006, \aj, 132, 1158

\bibitem[{{Ward}(2008)}]{VERITAS_1908_2008}
{Ward}, J.~E. 2008, in American Institute of Physics Conference Series, Vol.
  1085, American Institute of Physics Conference Series, ed. {F.~A.~Aharonian,
  W.~Hofmann, \& F.~Rieger}, 301--303

\bibitem[{{Watters} {et~al.}(2009){Watters}, {Romani}, {Weltevrede}, \&
  {Johnston}}]{Watters09}
{Watters}, K.~P., {Romani}, R.~W., {Weltevrede}, P., \& {Johnston}, S. 2009,
  \apj, 695, 1289

\bibitem[{{Yang} {et~al.}(2006){Yang}, {Zhang}, {Cai}, {Lu}, \&
  {Tan}}]{Yang2006}
{Yang}, J., {Zhang}, J.-L., {Cai}, Z.-Y., {Lu}, D.-R., \& {Tan}, Y.-H. 2006,
  Chinese Journal of Astronomy and Astrophysics, 6, 210

\end{thebibliography}
